# Fractal and multifractal analysis of PET/CT images of metastatic melanoma before and after treatment with ipilimumab

Christina-Marina Breki[1,2], Antonia Dimitrakopoulou-Strauss[3], Jessica Hassel[4], Theoharis Theoharis[2,5], Christos Sachpekidis[3,6], Leyun Pan[3], Astero Provata[1]

[1] Institute of Nanoscience and Nanotechnology, National Center for Scientific Research "Demokritos", Athens, Greece
[2] Department of Informatics & Telecommunications, National and Kapodistrian University of Athens, Athens, Greece
[3] Clinical Cooperation Unit Nuclear Medicine, German Cancer Research Center, Heidelberg, Germany
[4] National Center for Tumor Disease, Heidelberg, Germany and Department of Dermatology, University Hospital Heidelberg, Heidelberg, Germany
[5] Visual Computing Laboratory, Department of Computer and Information Science, Norwegian University of Science and Technology, Trondheim, Norway
[6] Department of Nuclear Medicine, Inselspital, University Hospital and University of Bern, Bern, Switzerland

correspondence address:

Antonia Dimitrakopoulou-Strauss, Prof. Dr.

Clinical Cooperation Unit Nuclear Medicine

German Cancer Research Center (DKFZ)

DE-69120 Heidelberg,Germany

Email: a.dimitrakopoulou-strauss@dkfz.de or ads@ads-lgs.de

## ABSTRACT

**Purpose:** PET/CT with F-18-Fluorodeoxyglucose (FDG) images of patients suffering from metastatic melanoma have been analysed using fractal and multifractal analysis to assess the impact of monoclonal antibody ipilimumab treatment with respect to therapy outcome.

**Methods:** 31 cases of patients suffering from metastatic melanoma have been scanned before and after treatment. We calculated the fractal and multifractal dimensions using the boxcounting method on the digitalised PET/CT images prior and after treatment to assess the therapeutic outcome. We modelled the spreading of malignant cells in the body via kinetic Monte Carlo simulations to address the dynamical evolution of the metastatic process and to predict the spatial distribution of malignant lesions.

**Results:**. Our analysis shows that the fractal dimensions which describe the tracer dispersion in the body decrease consistently with the deterioration of the patient's therapeutic outcome condition. In 20 out-of 24 cases the fractal analysis results match those of the medical records, while 7 cases are considered as special cases because the patients have non-tumour related medical conditions or side effects which affect the results. The decrease in the fractal dimensions with the deterioration of the patient conditions (in terms of disease progression) are attributed to the hierarchical localisation of the tracer which accumulates in the affected lesions and does not spread homogeneously throughout the body. Fractality emerges as a result of the migration patterns which the malignant cells follow for propagating within the body (circulatory system, lymphatic system). Analysis of the multifractal spectrum complements and supports the results of the fractal analysis. In the kinetic Monte

Carlo modelling of the metastatic process a small number of malignant cells diffuse throughout a fractal medium representing the blood circulatory network. Along their way the malignant cells engender random metastases (colonies) with a small probability and, as a result, fractal spatial distributions of the metastases are formed similar to the ones observed in the PET/CT images.

**Conclusions:** We show that the Monte Carlo generated spatial distribution of metastases changes with time approaching values close to the ones recorded in the metastatic patients. Thus, we propose that fractal and multifractal analysis has potential application in the quantification of the evaluation of PET/CT images to monitor the disease evolution as well as the response to different medical treatments.

**INTRODUCTION**

Many constituent networks of the human body present complicated geometric structures which derive from their functional needs, i.e. the geometry of the network serves its functionality. Common examples are: a) the circulatory system which is responsible for the efficient distribution of blood in the body, b) the respiratory system which coordinates the transport of oxygen in the body and the collection and release of carbon dioxide and c) the nervous system which controls the dynamic transmission of information via electrical signals in the brain and throughout the body [1]. These three major systems are complex structures with branching and sub-branching down in many hierarchical orders. Branching is considered as a very efficient and energy saving architecture to transfer matter (or information) from a central source (organ) to a large number of destinations. Because of branching the body systems are self-similar in many orders of magnitude and exhibit fractal scaling.

In particular, the blood circulatory system distributes the blood in the whole body starting from an organ of large size and reaching individual cells. To achieve this, the distribution network starts from a large artery leaving the heart and branches out in many levels. At each level, the number of branches increases while their diameter decreases in order to reach and feed all cells. This is a mathematical property of fractals in 3D which are constructed as space filling objects [2].

There has been a large number of attempts to quantify the value of the fractal dimensions of the cardiovascular system. Already in 1977, Mandelbrot predicted that the fractal dimension of an arterial tree should be smaller than **3** and he coined the value **2.7** as plausible in refs.[2] [3]. Huang et al. measured experimentally a value **$d_f$=2.71** for pulmonary arteries and **$d_f$=2.69** for pulmonary veins [4]. Helmberger et al reported fractal dimensions of the lung vessels in pulmonary hypertension patients as **$d_f$ = 2.34**, with **$d_f$** increasing to the value **2.37** for patients without pulmonary hypertension [5]. Gil-García et al, showed experimentally that the arterial pattern of the dog kidneys has fractal dimension **~2.7** [6].

The blood circulatory system often serves as a means of transportation for malignant cells to migrate through the body. As stem cancer cells migrate, they colonize distant areas producing metastases, which are the most frequent causes of death for patients with cancer. The molecular mechanisms of metastasis are still not well understood due to their biological complexity [7], but the migration through the blood circulatory system seems plausible since all body cells come in direct contact and exchange with blood. With this scenario the pattern of migration and colonization of malignant cells should be directly linked with the structure of the distributing blood network. It is then reasonable to search for fractality in the distribution of metastases in the human body, since the transportation network distributing the cancer cells is fractal. In many cases the evolution of cancer (melanoma in this case) can take a long time and it is a dynamical process. The final distribution is achieved only at the latest stages, while fractality changes during the various stages of the disease, as the spatial distribution of metastatic structures expands in the body.

One molecular imaging technique for the detection of primary tumours and metastases is Positron Emission Tomography (PET) using the radioactive biomarker F-18-Deoxyglucose (FDG). This technique allows the detection of viable tumour tissue and is used for staging and therapy monitoring of melanoma patients since several years [8][9][23][24][25]. In particular, the use of hybrid systems, like PET/CT and PET/MRI allows a better anatomic allocation of the PET findings and leads to an improvement of both diagnosis and treatment planning [10].

Metastatic melanoma has a poor diagnosis with a median overall survival of less that one year until recently [11]. The conventional treatment consisted in chemotherapy, radiotherapy, high-dose interleukin-2 and best supportive care. An improvement of the overall survival was achieved after the discovery of ipilimumab [22] [26]. Ipilimumab is a fully human, recombinant monoclonal antibody that activates the immune system by targeting cytotoxic T-lymphocyte-associated antigen 4 (CTLA-4). CTLA-4 is a type 1 transmembrane glycoprotein whose expression is primarily restricted to T cells [12]. It is a key negative regulator of T lymphocyte activation via antagonism of CD28-mediated co-stimulation. Melanoma, and tumours in general, can take advantage of this inhibitory mechanism used by the immune system to reduce T cell response. Therefore, the use of blocking antibodies against this inhibitory checkpoint, like ipilimumab, can enhance anti-tumour response.

Bearing these in mind, it is reasonable to use fractal and multifractal analysis as a quantification procedure for the evaluation of FDG images to monitor the disease evolution as well as the patient's response to different medical treatments. In the current study we demonstrate the use of fractal/multifractal analysis in detecting the response of **31** melanoma patients to treatment with ipilimumab monoclonal antibody, as will be explained in the sequel.

In the next section we describe the properties of the patients group and the fractal and multifractal methods used in the analysis of the spatial distribution of the metastatic lesions. In section 4 we present the main results of our study before and after treatment with the monoclonal antibody. Except from the results of the core group of patients we also present some specific cases which have non-tumor related findings. Also we present a minimal Kinetic Monte Carlo model which involves propagation of malignant cells through a fractal network causing occasional metastases and we compare the numerical results to those of the analysis of the PET/CT images. Finally, we recapitulate our main results and discuss open problems.

## 3. MATERIALS AND METHODS

### 3. 1 Patients

The study involved **31** patients (**P1**, **P2**, …, **P31**) who suffered from metastatic melanoma at various stages of the disease. Each patient was scanned once before treatment (**Study-I,** baseline scan) and in follow-up, during and after treatment with the monoclonal antibody ipilimumab. Follow-up scans were performed after two cycles of ipilimumab treatment (**Study-II**, first follow-up scan), and at the end of the four-cycle treatment (**Study-III**, second follow-up scan). Due to the tendency of melanoma to produce metastases throughout the body, whole-body scans were performed in all three studies of each patient.

Patients gave written informed consent to participate in the study and to have their medical records released. The study was approved by the Ethical Committee of the University of Heidelberg and the Federal Agency of Radiation Protection.

**3.2 Data acquisition**

Whole-body PET/CT studies were performed from the skull to the toes with an image duration of two minutes per bed position for the emission scans. A dedicated PET/CT system (Biograph mCT, S128, Siemens Co., Erlangen, Germany) with an axial field of view of 21.6 cm with TruePoint and TrueV, operated in a three-dimensional mode was used. A low-dose attenuation CT (120kV, 30 mA) was used for attenuation correction of the PET data and for image fusion. All PET images were attenuation-corrected and an image matrix of 400 x 400 pixels was used for iterative image reconstruction. Iterative image reconstruction was based on the Ordered Subset Expectation Maximization Algorithm (OSEM) with six iterations and twelve subsets. The reconstructed images were converted to Standardized Uptake Value (SUV) images based on the formula:

$$SUV = \frac{tissue\ consentration(B_q/g)}{[injected\ dose\ (B_q)/body\ weight(g)]} \quad (3.1)$$

SUVs were calculated for the lesions with the most intense tracer uptake and a functional tumor diameter larger than 1 cm, which was considered to be malignant. These lesions were re-evaluated in the follow-up studies.

As was explained in section 3.1, for each patient **Pi , i=1, …, 31,** three whole-body scans were obtained: the baseline scans and the two follow-up scans. Each one of these three sequences comprises about 400 images. The specifications of the PET images are given in Table 1.

It must be noted here that F-18-Deoxyglucose(FDG) also concentrates physiologically in several healthy organs where metabolic activity takes place. These organs are the heart, the brain, the liver, the urinary tract (due to tracer extraction) etc. Because it is not possible to systematically differentiate between non-malignant enhanced FDG activity (e.g. in inflammatory lesions) and abnormal activity in the tumorous lesions, a systematic error is introduced in the calculations of the fractal dimensions. This systematic error will be discussed further in the results section.

### 3.3. Data analysis

Evaluation of the PET/CT data was performed using a dedicated software (Pmod Technologies, Zuerich, Switzerland and Aycan Digitalsysteme, Würzburg, Germany). Visual analysis was performed by two nuclear medicine physicians evaluating the hypermetabolic areas on transaxial, coronal, and sagittal images which were considered to be malignant. The semi-quantitative (SUV) evaluation was based on irregular Volumes-of-Interest (VOIs), drawn with an isocontour based on a pseudosnake algorithm, placed over foci of increased 18F-FDG uptake in the selected areas. In general terms, foci presenting a significantly enhanced 18F-FDG uptake were considered indicative for melanoma after comparison with the morphology obtained by the CT images. Patient history was studied thoroughly in order to exclude possible causes of non-specific tracer accumulation and therefore minimize false-positive findings. Tumour response was based on the EORTC criteria. Details about the data evaluation have been previously published [8].

### 3.4 Fractal and Multifractal analysis

As discussed in the Introduction, melanoma is a fast evolving tumour which expands forming numerous lesions throughout the body. To quantify the spatial extension of the metastases, we use fractal analysis which takes into account only the positions of the lesions, while multifractal analysis accounts also for the size of the metastases in each affected area. The advantage of using fractal analysis to address this problem is that there exist a number of growth models / schemes which give rise to fractal structures. Depending on the outcome fractal dimension we can then point to the appropriate mechanisms, responsible for the spreading of the tumours.

From the many measures which are proposed for calculating fractality of the spreading, we use here the most classic one, the box-counting dimensions. This measure has the advantage of being easy to implement and it gives directly a means of differentiating between homogeneous and hierarchical arrangement of the lesions.

To implement the box-counting method, we divide the 3D PET/CT image of the human body in cubic boxes (3D pixels-voxels) of linear size **s**. The voxel linear size was chosen considering the pixel size and the image thickness (Table 1). According to the images specifications the pixel size is approximately **2mm X 2mm** and the image thickness is **4mm**. To be consistent in the three directions we construct cubic boxes with minimum length **$s_{min}$ = 4mm** up to **$s_{max}$ = 200 mm**.

For each value of box size **s** we calculate the number of cubic boxes **N(s)** which contain tumour cells (primary or metastatic). The presence of tumour cells in a given position is recorded by the presence of the tracer. By plotting **N(s)** as a function of **s** we extract the fractal dimension **$d_f$** fitting the data to the following formula:

$$N(s) = C s^{-d_f} \qquad (3.2)$$

where **C** is a constant, determined also by the above fit. Alternatively, we can plot **N(s)** as a function of **s** in a double logarithmic scale and the fractal dimension is calculated as the slope of this curve.

In these calculations a systematic error arises because the tracer accumulates not only in malignant regions but also in some healthy organs as well as in non-tumor related findings (e.g. inflammatory lesions). This error tends to increase the fractal dimension towards the value **3**, since the organs contribute substantial accumulations of tracer in 3D volumes.

While fractal analysis uses only the presence of the tracer in the particular lesion, multifractal analysis takes also into account the concentration of the tracer which is accounted for by the intensity **$p_i$** of the color in each voxel **i**. In this case we use the smallest box sizes available, because the multifractal spectrum requires the finest details of the spatial structure. For the present scans the smallest box size available is (**4mm x 4mm x 4mm**). The generalized dimensions **$D_q$** (or multifractal spectrum) are calculated as follows:

$$D_q = \lim_{s \to 0} \frac{1}{q-1} \frac{1}{\ln s} \sum_{i=1}^{N} p_i^q \qquad (3.3)$$

where **q** is the dimension index which takes real values **(-∞,..., 0, …, +∞), s = 4mm**, the index **i** runs over all boxes and **N** is the total number of boxes of size (**4mm x 4mm x 4mm**) covering the body. Note that the negative values of **q** highlight the smallest values of **$p_i$** (rare events), while the larger positive values of **$p_i$** are accentuated by the positive values of **q**.

**4. RESULTS**

**4.1 Typical Patients**

For each patient we calculated the box-counting dimensions according to formula 3.2, before medical treatment (**Study-I**, baseline scan), after two cycles of ipilimumab treatment (**Study-II**, first follow-up scan), and at the end of the four-cycle treatment (**Study-III**, second follow-up scan).

The general analysis and comparison between the fractal dimensions before and after treatment shows that the tracer has the tendency to spread in higher dimensionality in healthy subjects and also when the treatment is successful. Typical healthy subjects demonstrate fractal dimensions (in physiological distribution of the tracer) around **2.75.** The tracer tends to spread almost homogeneously throughout the body in healthy subjects, or concentrates in organs where biochemical activity takes place. In these cases the spreading shows fractal dimension almost equal (very close) to **3**. Instead, when lesions are spread throughout the body the fractal dimensions decrease because the tracer tends to concentrate in a subspace of lower dimensions, where the metastases are active. Using the patients' medical records as reference, we see that when the treatment with ipilimumab is successful the fractal dimension increases, when the patient's condition is stable $d_f$ does not change drastically, and when the patient's condition deteriorates $d_f$ decreases further.

A representative plot for patient **P3**, is shown in Fig. 4.1. With the red line we represent the analysis of Study-I, when the metastatic melanoma was diagnosed, with the blue line is the analysis after two cycles of ipilimumab treatment (Study-II) and with the green line the analysis after four cycles of treatment (Study-III). As indicated by the medical records, the patient at the baseline Study-I showed low fractal dimensions $d_f(I)=2.46$ (metastatic invasion). After two cycles we have an improvement of the patient's condition and $d_f(II)$ increased to **2.64**, and after four cycles $d_f(III)$ did not change substantially, showing a stable phase of the disease. In the same figure, the line corresponding to the analysis of a healthy subject is plotted, for comparison.

The results of the multifractal spectrum verify further the fractal analysis. For the same patient the generalized dimensions are presented in Fig. 4.2. Again, the red line which corresponds to the initial diagnosis of the disease (Study-I) is lower than all other lines. The blue line which records the analysis after two cycles of medical treatment (Study-II) and the green line, after four medical treatments (Study-III) are at the same level higher than the baseline spectrum. This indicates improvement in the patient conditions and elimination of malignant lesions. The black line which denotes the multifractal spectrum of a healthy subject is at the same level as the Studies II and III.

Another representative plot from patient **P12**, whose condition according the medical records deteriorates during therapy, is shown in Fig 4.3. In this case, the $d_f$ value drops successively among the different studies. On Study-I the fractal dimension showed the highest value $d_f(I)=2.63$. As the disease progressed a decrease in $d_f$ values is shown with $d_f(II)=2.59$ and $d_f(III)=2.56$ respectively. The above confirms the hypothesis that in the presence of developing malignant lesions the tracer tends to accumulate in them, showing hierarchical expansion of low dimensionality. Figure 4.4 presents the multifractal spectrum for the same patient. The curve shapes corroborate the results of the fractal analysis showing a similar decreasing tendency.

The results for all patients are recorded on Table **2.** In the 1$^{st}$ column the patient index is given, as **P1**, **P2**, … **P31**. In the 2$^{th}$ and 3$^{th}$ column the patient's age and sex are recorded. In the 4$^{th}$ column the three therapy stages are set, while in the 5$^{th}$ column the medical records are reported based on clinically used response criteria for therapy response evaluation. The medical records characterize the progression of the disease between studies I and II **(Early)**, studies II and III **(Late)** and studies I and III **(Final)**. The patients condition is characterized as **Partial Remission (PR)** if the patient improves after treatment, as **Stable Disease (SD)** if the patient's condition is stable, as **Progressive Disease (PD)** if the patient deteriorates and as **Mixed Response (MR)** in case of inability to evaluate properly patient's condition. The 6$^{th}$ column records the fractal dimension characteristic of the particular study. In the 7th column the normalized integral difference of the multifractal spectrum, Average Multifractal Index $D_{MF}$ , between studies I and II (Early), II and III (Late) and I and III (Final) is reported. Formula 4.1 is used for the calculation of $D_{MF}$ between the early and the late stage, and similarly between the late and final stages,

$$D_{MF_{(II,I)}} = \frac{1}{2L+1} \sum_{q=-L}^{L} \left( D_q(II) - D_q(I) \right) \qquad (4.1)$$

where **L** (**-L**) is the highest (lowest) order of index **q**. In the last column the patient is characterized as **Standard/Typical (Typ)** if he/she only suffers from metastatic melanoma and the comparison of the analysis results and the medical records follow the typical pattern as explained in section **4.1**. The patient is characterized as **Non-Typical** (**non-Typ**) if he/she has a non-tumor related lesion or if he/she displays a non-standard mechanism of tracer spreading which makes the connection between the analysis and the medical records inconsistent. Patients characterized as non-Typ are individually discussed in the next section, **4.2.**

**4.2 Unspecific tracer uptake not related to melanoma disease**

In this section we present two cases of non-typical patients, both of which show pre-existing medical findings, not directly related to the melanoma disease. In both cases there is an uptake of FDG in non-tumorous areas and thus the results of the fractal/multifractal analysis are incompatible with the medical records.

*4.2.1 The case of colitis*

Patient **P15** who suffers from metastatic melanoma presents one metastatic lesion in the 6th thoracic vertebra, which is clearly delineated in the maximum intensity projection images in both follow-up studies (see Fig. 4.5). The same patient suffers also from fast evolving colitis, which is a side-effect of ipilimumab therapy and this is evident from the high uptake of FDG in the colon as appeared on Study-III. Furthermore, there is tracer retention in both ureters in the third study, which is also a finding not-related to the melanoma disease. Thus the distribution of FDG is divided between healthy organs, metastatic lesions and the colon area. This introduces a large error in the calculations and as a result the fractal dimensions increase in the Study II and decrease in the Study III, while the patient's condition remains practically steady.

*4.2.2 The case of unspecific FDG uptake (thrombophlebitis)*

Patient **P31** in the stage of treatment (Study-II) presents an increase in FDG uptake in the left leg which disappears in the Study-III. This uptake is not related to fast progressing melanoma but to an inflammatory vessel disease (see Fig. 4.6). Due to this higher dispersion of the FDG biomarker the fractal dimension decreases in Study-II, as if the patient was deteriorating (developing further lesions) while the medical records indicate that the patient's condition is stable. Similar, hypermetabolic lesions not directly related to melanoma (unspecific findings) altering the results of the fractal/multifractal analysis are met in **7** out of **31** patients.

The results of this section demonstrate that the fractal/multifractal analysis predicts correctly the evolution of metastatic melanoma in **83.3%** (20 in 24) of the cases, giving results compatible with the medical records. In **7** cases, combined effect of melanoma and other non-specific findings (e.g. inflammatory lesions) gives results non-compatible with medical records. We stress again that the fractal/multifractal analysis of the metastatic melanoma spreading must be always complemented with medical assessment in order to exclude regions with false positive FDG uptake not related to the disease.

## 5. DISCUSSION

In this section we first present the main results produced by a Kinetic Monte Carlo **(KMC)** model which mimics the transportation of malignant cells in the body. The model assumes biologically motivated processes such as diffusion of malignant cells through the blood (fractal) circulatory system and aggregation/colonization in the form of metastases at first contact of the diffusing cells with the surrounding healthy tissue. The results of this model are compared with the results of the PET/CT fractal analysis in section **5.2** while the limitations of this method and algorithms are discussed in **5.3**.

### 5.1 Kinetic Monte Carlo simulations and results

Previous attempts to model the growth of tumours include both continuous and discrete approaches; for a review see ref. [13]. Most of these models deal with the growth of a single lesion as a result of the increase of the number of malignant cells locally. The propagation of tumour cells causing metastases has been studied using mainly continuous diffusion models [14],[15],[16]. In the current study we employ a stochastic microscopic approach [18],[19],[20], which allows the malignant cells to move randomly through the blood circulatory system. To model the metastasis process of melanoma we allow a small number of malignant cells to diffuse through the blood circulatory system and to cause occasional metastases in the adjacent healthy tissue or organs. The details of the kinetic Monte Carlo Algorithm and the spatial properties (fractal dimensions) of the distribution of the metastases are presented in detail in the Appendix. The main results of the KMC model is that as the malignant cells diffuse through the blood network colonizing at first contact (with given, small probability), the colonies at the beginning of the process seem to be scattered randomly in the body. As time evolves, and more and more metastases occur, the metastatic topology will develop following the shape of the accommodating blood circulatory (or other hosting) network. This is clearly shown in the time evolution of the fractal dimensions, which as time evolves approach the limiting set with **$d_f$=2.7,** characteristic of biological networks (see Appendix). The number of metastases in the KMC model is shown to follow a mixed mathematical increasing expression containing both exponential and power law parts, characteristic of fractal growth phenomena [17]. This mathematical formula can be used as predictor for the extension of the "volume" of the metastatic lesions with time, provided that more frequent scannings of the patients are available in order to have enough data to accurately fix the values of the constants **A, α** and **β** in formula A.1.

## 5.2 Fractal and multifractal properties of the PET/CT images

We have used the methods of fractal and multifractal analysis to quantify the spatial extension of the malignant lesions in patients with metastatic melanoma, before and after treatment with ipilimumab. We analysed the PET/CT data from **31** patients at various stages of the disease. The fractal dimensions show low, decreasing values in patients with progressive disease, demonstrating the tendency of the tracer to concentrate in the sub-region of the body where metabolic activity takes place. However, in cases of successful medical treatment the lesions cease to attract the tracer and the fractal dimensions increase, showing the tendency of the tracer to diffuse in all parts of the body. In **20** out of **24** cases (**83.3%**) the results of the fractal and multifractal analysis were consistent with the medical records. For the remaining 7 patients non-tumour related medical conditions led to abnormal tracer uptake, which gave false quantitative results. Two of these cases were demonstrated in detail in sect. **4.2** as common atypical cases.

To understand the quantitative results we have devised a stochastic kinetic model, based on the random drift of malignant cells through the blood circulatory system, which colonise randomly the surrounding tissue. The KMC model produces spatial distribution of metastases statistically similar to the ones observed in melanoma patients.

**5.3 Open Problem and Limitations**

One limitation of the present study is the selectivity of the algorithm used for the calculation of the fractal and multifractal dimensions. An important improvement to the algorithm would be the ability to exclude healthy organs where the tracer physiologically accumulates, as well as areas which show an unspecific non-tumour related FDG uptake. This became evident during the data evaluation and after the comparison between the results of the fractal analysis and the medical records. This problem requires the development of new, smart algorithms, able to mask areas with non-tumour related tracer uptake.

 Other nonlinear measures such as the spatial correlations, higher order moments of the spatial distribution and clustering coefficients can be used to characterise the metastatic potential of the tumours and can improve the predictions of therapy outcome. These measures can be easily incorporated in the existing algorithms.

**6. CONCLUSIONS**

In this study we present a new method based on the calculations of fractal and multifractal dimensions of PET/CT images with FDG for monitoring of the therapeutic effect of ipilimumab in patients with metastatic melanoma. The results show enough evidence that the fractal and multifractal analysis have the potential to serve as biomarkers, for therapy monitoring and for following the evolution of metastases in the body. This is a robust, operator independent method which is based on static images, can be easily implemented and demonstrated a correct classification rate of 83.3% in this study. It can be used as an additional quantitative parameter for the assessment of the therapeutic outcome in clinical routine. Additional studies over a larger number of patients are needed to verify the impact of this method, as well as refinement of the developed algorithms to improve their predictive value.

**COMPLIANCE WITH ETHICAL STANDARDS**

**Conflict of Interest:** A. P. and C. M. B. receive traveling support by the European Union (European Social Fund – ESF) and Greek national funds through the Operational Program "Education and Lifelong Learning" of the National Strategic Reference Framework (NSRF) - Research Funding Program: THALES. Investing in knowledge society through the European Social Fund.

ADS received support from the German Cancer Society (Deutsche Krebshilfe) for the project entitled: "Therapy monitoring of ipilimumab based on quantification of the F-18-Deoxyglucose (FDG) kinetics with the 4D positron emission tomography/computed tomography (dPET-CT) in patients with melanoma (stage 4)".

All other authors have nothing to declare.

**Ethical approval:** All procedures performed in studies involving human participants were in accordance with the ethical standards of the institutional and/or national research committee and with the 1964 Helsinki declaration and its later amendments or comparable ethical standards. The study was approved from the Ethical Committee of the University of Heidelberg and the Federal Agency of Radiation Protection.

**Informed consent:** Informed consent was obtained from all individual participants included in the study.

**APPENDIX: KINETIC MONTE CARLO MODELING**

According to earlier studies, the blood circulatory system has a branching structure whose geometry is fractal [3],[6]. As recapitulated in the Introduction, the fractal dimensions recorded in the literature is of the order of **2.6 - 2.7.** For this study, the blood circulatory network was constructed as a deterministic fractal embedded in **3D** space representing the human body. The system (body) size was **L x L x L= 81 x 81 x 81** sites (**3D** representation) and the deterministic fractal network corresponding to the circulatory system was constructed iteratively within this 3D space. The fractal network spans the body and its fractal dimension $d_f$ was chosen $d_f^c = \ln(18) / \ln(3) =$ **2.69** to be similar to the value reported in the literature. Once the positions of the sites belonging to the circulatory system are assigned, a small number (**n=10**) of malignant cells are released to diffuse [18],[21]. Their motion is restricted only within the areas covered by the circulatory network, i.e. they do not enter the areas outside the circulatory system, but occasionally, with very small probability (**p**) they infect the tissue (or organs) adjacent to the circulatory system. As time passes, the number of tumor lesions increases and they span more and more areas around the body, all of them adjacent to the circulatory system.

To find the characteristics of the spatial distribution of the metastatic lesions produced by the KMC model, we use the same algorithm (box counting method) as in the study of the PET/CT images, i.e. we divide the **81 x 81 x 81** sites in cubic boxes of size **s = 1,2, .... 20**, and we calculate the number of boxes **N(s)** which contain at least one infected site. In a double logarithmic scale, the tangent of the number of boxes **N(s)** versus the box size **s** expresses the fractal dimensions **$d_f$(sim)**. The results of a typical simulation are shown in Fig. A.1. In this simulation each malignant cell diffuses for time **T=100 x $L^3$** where **L** is the linear size of the body (in voxels). At every elementary time step they diffuse randomly into nearest neighbouring sites (provided that they belong to the circulatory system) and with probability **p=0.001** they cause metastases in the nearby tissues or organs.

The number of infected regions at this time shows a typical power law with exponent **$d_f$(sim)=2.76,** close to the one of the circulatory system. In earlier times the metastases distribution is more sparse and the evolution of the fractal dimension with time is shown on Fig. A.2. This figure shows that as time increases the fractal dimension of the metastatic lesions approaches closely the one of the circulatory system. In fact, it tends to be a little higher. This is explained because the lesions are located all around the blood network covering a larger area than the network itself. Consequently, the corresponding fractal dimension tends to be relatively higher, in the limit of large times. We believe that the recorded PET/CT scans represent earlier, transient times before the malignant cells have the time to infect all the maximum body "volume" around the circulatory system.

Fractal laws have been also demonstrated in the time activity curve of FDG in different tumours [27]. A similar tendency is demonstrated by the number of lesions (metastases) **$N_m$** as a function of time shown in Fig. A.3. **$N_m$** increases at the beginning while after the transient it approaches a constant value, since most regions around the circulatory system have been infected and secondary infection of the same site is not allowed in the simulations. In the same figure the simulation results are approximated using a nonlinear curve fit of the type [21]:

$$N_m(t) = At^\alpha \exp(-\beta t) \qquad (A.1)$$

This formula takes into account the power law (fractal) nature of the problem, expressed by the exponent **α** and the leveling up of the curve which is attained by the exponential decay term **exp(-βt)**. The nonlinear curve fit gives parameter values **α= 0.808** and **β= -0.0045**, with correlation coefficient **0.999697**.

The minimal KMC model is a mechanistic scheme for the spreading of metastases in the human body. It accounts for the fractal spreading as calculated from the PET/CT images and has a similar fractal dimension. This model needs to be further explored in order to account for more precise metastasis characteristics. A first improvement would be to choose a spatial extension similar to the one used in PET/CT (eg. **400 x 400 x 400** depending on the patient height) rather than the simple cubic (**81 x 81 x 81**) used here. Instead of using a deterministic fractal we could mimic the circulatory network itself, drawing from medical images. We can also vary the number of malignant cells which circulate in the blood vessels and make their motion more realistic, rather than classical diffusion. It is also possible to arrange the malignant cells to start from a single tumorous area (e.g. the primary tumour) rather than spreading them randomly in the circulatory network. These additions to the KMC model could lead to even closer predictions of the spatial distribution of the lesions in metastatic melanoma.

**REFERENCES**

1. West BJ, Fractal Physiology and Chaos in Medicine, 2$^{nd}$ edition, World Scientific, 2013.

2. Mandelbrot B, Fractal Geometry of Nature. W. H. Freeman and Company, 1$^{st}$ edition, 1982.

3. Weibel ER, Fractal Structures in Biological Systems, in Fractals in Biology and Medicine, Vol. 5, Losa G, Merlini D, Nonnenmacher TF, Weibel ER (edts)., pages 3-18.

4. Huang W, Yen RT, McLaurine M, Bledsoe G, Morphometry of the human pulmonary vasculature, Journal of Applied Physiology, 1996; 81: 2123-2133.

5. Helmberger M, Pienn M, Urschler M, Kullnig P, Stollberger R, Kovacs G, Olschewski A, Olschewski H, Bálint Z, Quantification of tortuosity and fractal dimension of the lung vessels in pulmonary hypertension patients, PLoS One, 2014; 9:e87515.

6. Gil-García J, Gimeno-Domínguez M. Murillo-Ferroll NL, The arterial pattern and fractal dimension of the dog kidney, Histology and Histopathology 1992; 7: 563-574.

7. Yamaguchi H, Wyckoff J, Condeelis J, Cell migration in tumor, Current Opinion in Cell Biology, 2005; 17: 559-564.

8. Sachpekidis C, Larribere L, Pan L, Haberkorn U, Dimitrakopoulou-Strauss A, Hassel JC. Predictive value of early 18F-FDG PET/CT studies for treatment response evaluation to ipilimumab in metastatic melanoma: preliminary results of an ongoing study. Eur J Nucl Med Mol Imaging 2015;42:386-396.

9. Dimitrakopoulou-Strauss A, Strauss LG, Burger C. Quantitative PET studies in pretreated melanoma patients: a comparison of 6-(18F)fluoro-L-dopa with 18F-FDG and 15-O-water using compartment and noncompartment analysis. J Nucl Med 2001;42:248-256.

10. Dimitrakopoulou-Strauss A. PET-based molecular imaging in personalized oncology; potential of the assessment of therapeutic outcome. Future Oncol 2015;11:1083-1091.

11. Hodi FS, O'Day SJ, McDermott DF, Weber RW, Sosman JA, et al. Improved survival with ipilimumab in patients with metastatic melanoma. N Engl J Med 2010;368:711-723.

12. Brunet JF, Denizot F, Luciani MF, Roux-Dosseto M, Suzan M, et al. A new member of the immunoglobulin superfamily-CTLA-4. Nature 1987;32

13. Lowengrub J S, Frieboes H B , Jin F, Chuang Y-L, Li X, Macklin P, Wise SM, and Cristini V. Nonlinear modelling of cancer: bridging the gap between cells and tumour, Nonlinearity. 2010; 23(1): R1–R9.

14. Cruywagen GC, Woodward DE, Tracqui P, Bartoo GT, Murray JD, Alvord EC., Jr. The modeling of diffusive tumors. J. Biol. Systems. 1995; 3: 937–945.

15. Szymanska Z, Rodrigo CM, Lachowicz M, Chaplain MAJ. Mathematical modeling of cancer invasion of tissue: the role and effect of nonlocal interactions. Math. Models Methods Appl. Sci. 2009;19:257–281.

16. Ambrosi D, Duperray A, Peschetola V, Verdier C. Traction patterns in tumour cells, J Math Biol. 2009; 58(1-2):163-81.

17. Vicsek T., Fractal Growth Phenomena, World Scientific Publishing, Singapore, 1989.

18. Witten TA, Sander LM, Diffusion-Limited Aggregation, a Kinetic Critical Phenomenon, Phys. Rev. Lett. 1981; 47: 1400-1403.

19. Meakin P, Diffusion-controlled cluster formation in 2-6-dimensional space, Phys. Rev. A 1983; 27: 1495-1507.

20. Meakin P, Progress in DLA research, Physica D 1995; 86:104-112.

21. Meakin P, Fractals, scaling and growth far from equilibrium, Cambridge University Press, ISBN 0 521 45253 8, 1998.

22. Hodi FS, O'Day SJ, McDermott DF, Weber RW, Sosman JA, et al. Improved survival with ipilimumab in patients with metastatic melanoma. *N Engl J Med* 2010;368:711-723.

23. Holder WD Jr, White RL Jr, Zuger JH, Easton EJ Jr, Greene FL. Effectiveness of positron emission tomography for the detection of melanoma metastases. Ann Surg. 1998;227:764-769; discussion 769-771.

24. Mijnhout GS, Hoekstra OS, van Tulder MW et al. Systematic review of the diagnostic accuracy of 18F-fluorodeoxyglucose positron emission tomography in melanoma patients. *Cancer*. 2001;91:1530-1542.

25. Strauss LG, Conti PS. The applications of PET in clinical oncology. J Nucl Med. 1991;32:623-648

26. Weber JS, O'Day S, Urba W, Powderly J, Nichol G, et al. Phase I/II study of ipilimumab for patients with metastatic melanoma. J. Clin. Oncol. 2008;26:5950-5956.

27. Dimitrakopoulou-Strauss A, Strauss LG, Mikolajczyk K, Burger C, Lehnert T, Bernd L, Ewerbeck V. On the Fractal Nature of Dynamic Positron Emission Tomography (PET) Studies. World J. Nucl. Med. 2003;2:306-313.

**Tables:**

*Table 1:* *General features of PET images*

| Specifications of PET Images | |
|---|---|
| Image size: | 400 x 400 pixels |
| Pixel size: | 2.03642 x 2.03642 mm$^2$ |
| Slice Thickness: | 4 mm |
| Photometric Interpretation: | MONOCHROME2 |
| File Type: | .dcm (DICOM) |

*Table 2:* *Analysis Results for all Patients.*

| Patient no. | Age (years) | SEX (M/F) | Study | Treatment Response | Fractal Dimension | Average Multifractal Index | Matching Results |
|---|---|---|---|---|---|---|---|
| P1 | 67 | M | Early | * | 2.487 | -0.064 | Typ |
|  |  |  | Late | * | 2.509 | -0.030 |  |
|  |  |  | Final | PR | 2.525 | -0.095 |  |
| P2 | 48 | F | Early | PD(slow) | 2.611 | +0.043 | Typ |
|  |  |  | Late | MR | 2.588 | -0.029 |  |
|  |  |  | Final | MR | 2.623 | +0.013 |  |
| P3 | 56 | M | Early | * | 2.465 | -0.119 | Typ |
|  |  |  | Late | * | 2.644 | -0.015 |  |
|  |  |  | Final | PR | 2.643 | -0.135 |  |
| P4 | 62 | M | Early | PD | 2.688 | -0.108 | Typ |
|  |  |  | Late | MR | 2.743 | +0.003 |  |
|  |  |  | Final | PD | 2.726 | -0.104 |  |
| P5 | 37 | F | Early | PD (slow) | 2.512 | +0.181 | Typ |
|  |  |  | Late | PD (slow) | 2.395 | -0.308 |  |
|  |  |  | Final | PD (slow) | 2.602 | -0.127 |  |
| P6 | 55 | M | Early | PD | 2.634 | -0.024 | Typ |
|  |  |  | Late | PD | 2.673 | +0.174 |  |
|  |  |  | Final | PD | 2.511 | +0.150 |  |
| P7 | 55 | M | Early | PD | 2.65 | +0.077 | Typ |
|  |  |  | Late | SD | 2.629 | -0.024 |  |
|  |  |  | Final | PD | 2.624 | +0.053 |  |

(Table 2 continued)

| Patient no. | Age (years) | SEX (M/F) | Stage | Treatment Response | Fractal Dimension | Average Multifractal Index | Matching Results |
|---|---|---|---|---|---|---|---|
| P8 | 74 | M | Early | PD(slow) | 2.622 | +0.01 | Typ |
|  |  |  | Late | SD | 2.629 | +0.11 |  |
|  |  |  | Final | PD | 2.529 | +0.13 |  |
| P9 | 60 | F | Early | PD | 2.618 | +0.016 | Non-Typ |
|  |  |  | Late | PD | 2.603 | -0.14 |  |
|  |  |  | Final | PD | 2.655 | -0.124 |  |
| P10 | 67 | M | Early | SD | 2.589 | -0.065 | Non-Typ |
|  |  |  | Late | PD | 2.633 | -0.044 |  |
|  |  |  | Final | PD | 2.652 | -0.11 |  |
| P11 | 56 | M | Early | MR | 2.453 | -0.369 | Typ |
|  |  |  | Late | PD(slow) | 2.719 | +0.455 |  |
|  |  |  | Final | MR | 2.4 | +0.085 |  |
| P12 | 55 | M | Early | PD(slow) | 2.637 | +0.044 | Typ |
|  |  |  | Late | PD | 2.591 | +0.03 |  |
|  |  |  | Final | PD | 2.56 | +0.075 |  |
| P13 | 71 | M | Early | SD | 2.653 | -0.113 | Typ |
|  |  |  | Late | SD | 2.676 | +0.07 |  |
|  |  |  | Final | PR | 2.683 | -0.043 |  |
| P14 | 36 | M | Early | PD | 2.616 | +0.053 | Typ |
|  |  |  | Late | PD | 2.613 | +0.106 |  |
|  |  |  | Final | PD | 2.523 | +0.159 |  |

(Table 2 continued)

| Patient no. | Age (years) | SEX (M/F) | Stage | Treatment Response | Fractal Dimension | Average Multifractal Index | Matching Results |
|---|---|---|---|---|---|---|---|
| P15 | 55 | M | Early | SD | 2.593 | -0.264 | Non-Typ |
|  |  |  | Late | SD | 2.707 | +0.281 |  |
|  |  |  | Final | SD | 2.575 | 0.017 |  |
| P16 | 73 | M | Early | SD | 2.61 | -0.075 | Typ |
|  |  |  | Late | PD(slow) | 2.676 | +0.063 |  |
|  |  |  | Final | PD(slow) | 2.618 | -0.012 |  |
| P17 | 61 | F | Early | PD(slow) | 2.698 | +0.001 | Typ |
|  |  |  | Late | PR | 2.697 | -0.033 |  |
|  |  |  | Final | PR | 2.705 | -0.031 |  |
| P18 | 71 | M | Early | PD | 2.697 | -0.031 | Non-Typ |
|  |  |  | Late | PD | 2.684 | -0.047 |  |
|  |  |  | Final | PD | 2.713 | -0.078 |  |
| P19 | 39 | M | Early | PD(slow) | 2.611 | -0.036 | Typ |
|  |  |  | Late | SD | 2.629 | +0.088 |  |
|  |  |  | Final | PD(slow) | 2.572 | +0.052 |  |
| P20 | 70 | M | Early | PD | 2.663 | +0.045 | Typ |
|  |  |  | Late | PD | 2.659 | -0.021 |  |
|  |  |  | Final | PD | 2.621 | +0.024 |  |
| P21 | 49 | M | Early | MR | 2.533 | +0.012 | Non-Typ |
|  |  |  | Late | PD | 2.518 | -0.036 |  |
|  |  |  | Final | PD | 2.556 | -0.023 |  |

(Table 2 continued)

| Patient no. | Age (years) | SEX (M/F) | Stage | Treatment Response | Fractal Dimension | Average Multifractal Index | Matching Results |
|---|---|---|---|---|---|---|---|
| P22 | 66 | M | Early | SD | 2.572 | -0.037 | Typ |
|  |  |  | Late | SD | 2.561 | +0.031 |  |
|  |  |  | Final | SD | 2.582 | -0.005 |  |
| P23 | 75 | M | Early | PR | 2.68 | +0.09 | Non-Typ |
|  |  |  | Late | PD | 2.609 | +0.011 |  |
|  |  |  | Final | PD | 2.651 | +0.101 |  |
| P24 | 67 | M | Early | SD | 2.687 | +0.03 | Non-Typ |
|  |  |  | Late | PD(slow) | 2.67 | -0.071 |  |
|  |  |  | Final | PD(slow) | 2.741 | -0.04 |  |
| P25 | 65 | F | Early | MR | 2.48 | -0.319 | Typ |
|  |  |  | Late | MR | 2.699 | +0.151 |  |
|  |  |  | Final | MR | 2.614 | -0.168 |  |
| P26 | 73 | F | Early | PR | 2.65 | -0.081 | Non-Typ |
|  |  |  | Late | PD | 2.708 | -0.07 |  |
|  |  |  | Final | PD | 2.714 | -0.151 |  |
| P27 | 62 | M | Early | PD | 2.602 | +0.021 | Non-Typ |
|  |  |  | Late | PR | 2.599 | +0.047 |  |
|  |  |  | Final | PR | 2.576 | +0.069 |  |
| P28 | 54 | F | Early | PR | 2.699 | +0.129 | Typ |
|  |  |  | Late | PR | 2.655 | -0.187 |  |
|  |  |  | Final | PR | 2.716 | -0.058 |  |

| Patient no. | Age (years) | SEX (M/F) | Stage | Treatment Response | Fractal Dimension | Average Multifractal Index | Matching Results |
|---|---|---|---|---|---|---|---|
| P29 | 61 | F | Early | PR | 2.622 | -0.019 | Non-Typ |
|  |  |  | Late | PR | 2.632 | -0.032 |  |
|  |  |  | Final | PR | 2.651 | -0.052 |  |
| P30 | 53 | M | Early | SD | 2.544 | +0.051 | Typ |
|  |  |  | Late | SD | 2.534 | -0.203 |  |
|  |  |  | Final | SD | 2.676 | -0.152 |  |
| P31 | 78 | F | Early | SD | 2.588 | +0.046 | Non-Typ |
|  |  |  | Late | SD | 2.554 | -0.165 |  |
|  |  |  | Final | SD | 2.633 | -0.118 |  |
| Healthy Subject | 73 | M | - | - | 2.723 | - | - |

*no medical records

**PR** - Partial Remission, **SD -** Stable Disease, **PD -** Progressive Disease, **MR -** Mixed Response

**Early Stage** :Study-I → Study-II, **Late Stage**:Study-II → Study-III, **Final Stage**:Study-I →Study-III

**Typ** - correspondence between analysis results and medical records, **non-Typ** - no correspondence between analysis results and medical records

**Figure Legends:**

**Fig. 4.1.** The number of boxes **N(s)** as a function of the box size **s** for patient **P3**. His condition improves after medical treatment with ipilimumab.

**Fig. 4.2.** The generalized dimensions **Dq** as a function of the index **q** for patient **P3**. The patient's condition improves after medical treatment with ipilimumab.

**Fig. 4.3.** The number of boxes **N(s)** as a function of the box size **s** for patient **P12**. His condition deteriorates during medical treatment with ipilimumab.

**Fig. 4.4**. The generalized dimensions **$D_q$** as a function of the index **q** for patient **P12**. The patient's condition deteriorates during medical treatment with ipilimumab.

**Fig.4.5.** PET/CT images of patient **P24** suffering from colitis. The patient presents high uptake of the FDG biomarker not only due to the melanoma metastases visible in his/her left leg but also due to the presence of colitis, not related to the melanoma.

**Fig. 4.6.** PET/CT images of patient **P31** at the three stages. The patient presents high unusual uptake of the FDG biomarker in his/her left leg at the **2$^{nd}$ stage** of treatment, not related to the melanoma.

**Fig. A.1.** The number of boxes containing metastases as a function of the box size **s**. The linear system size is **L=81,** the probability of infection is **p=0.001,** the infecting seeds are **n=10** and the time of the simulation is **T=100 x $L^3$**.

**Fig. A.2.** The temporal evolution of **$d_f$(sim)**.

**Fig. A.3.** Evolution of the metastatic lesions with time. The crosses represent the simulation results and the red dashed line the nonlinear curve fit.

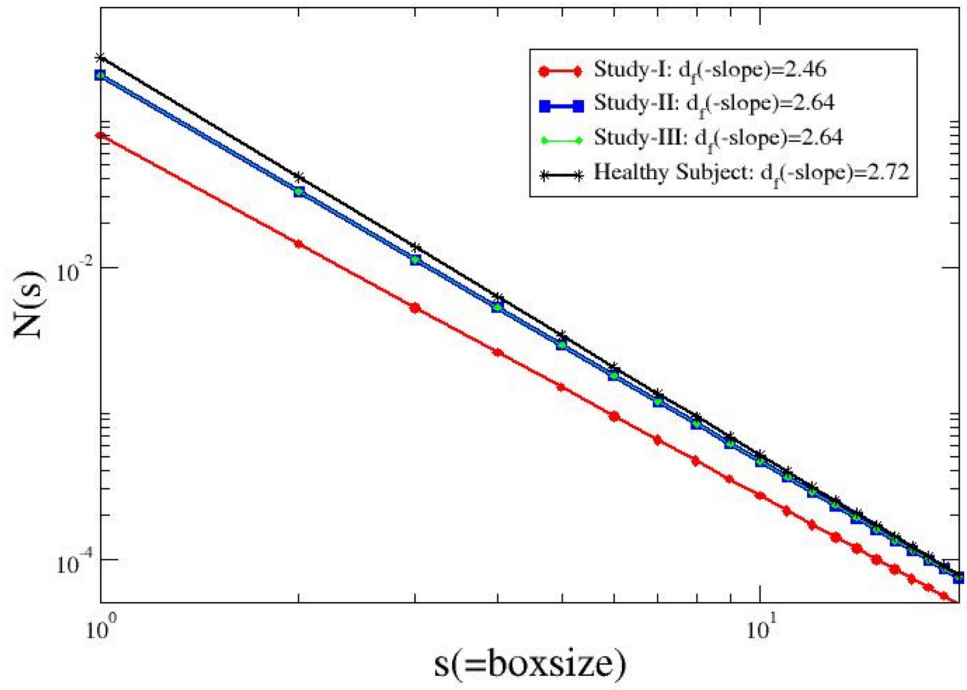

*Fig. 4.1.*

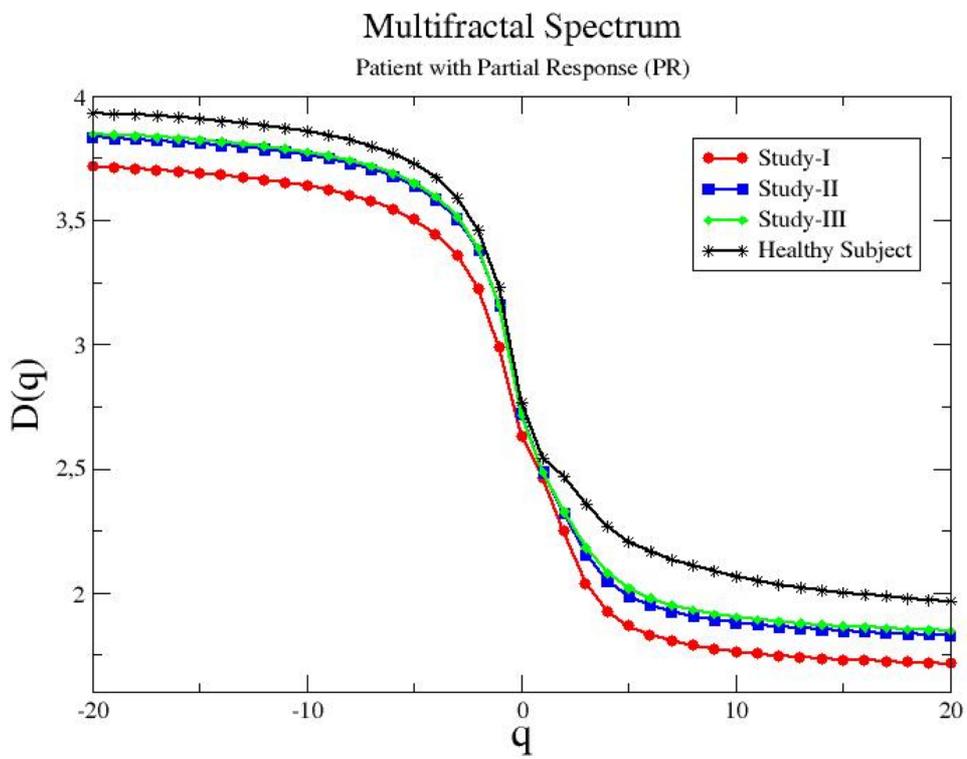

*Fig. 4.2.*

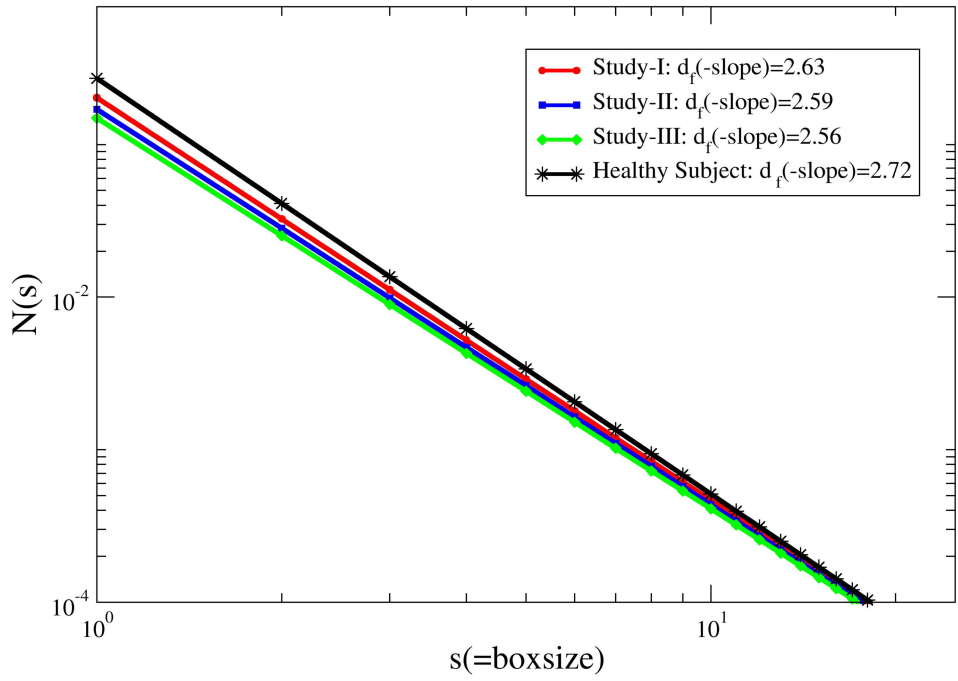

Fig. 4.3.

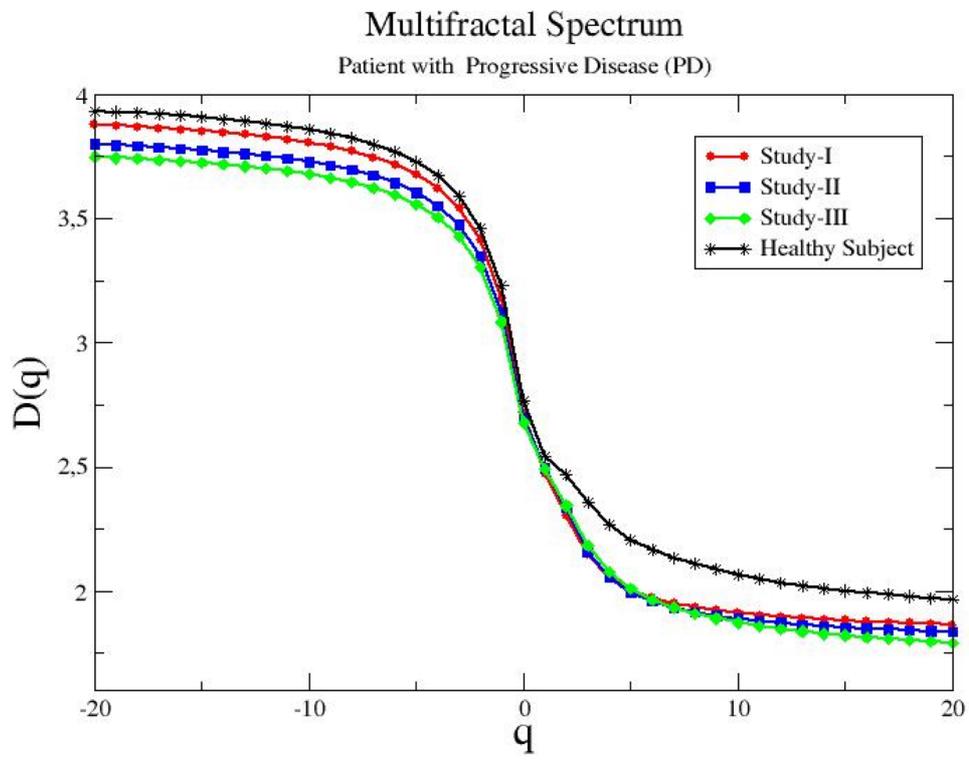

*Fig. 4.4.*

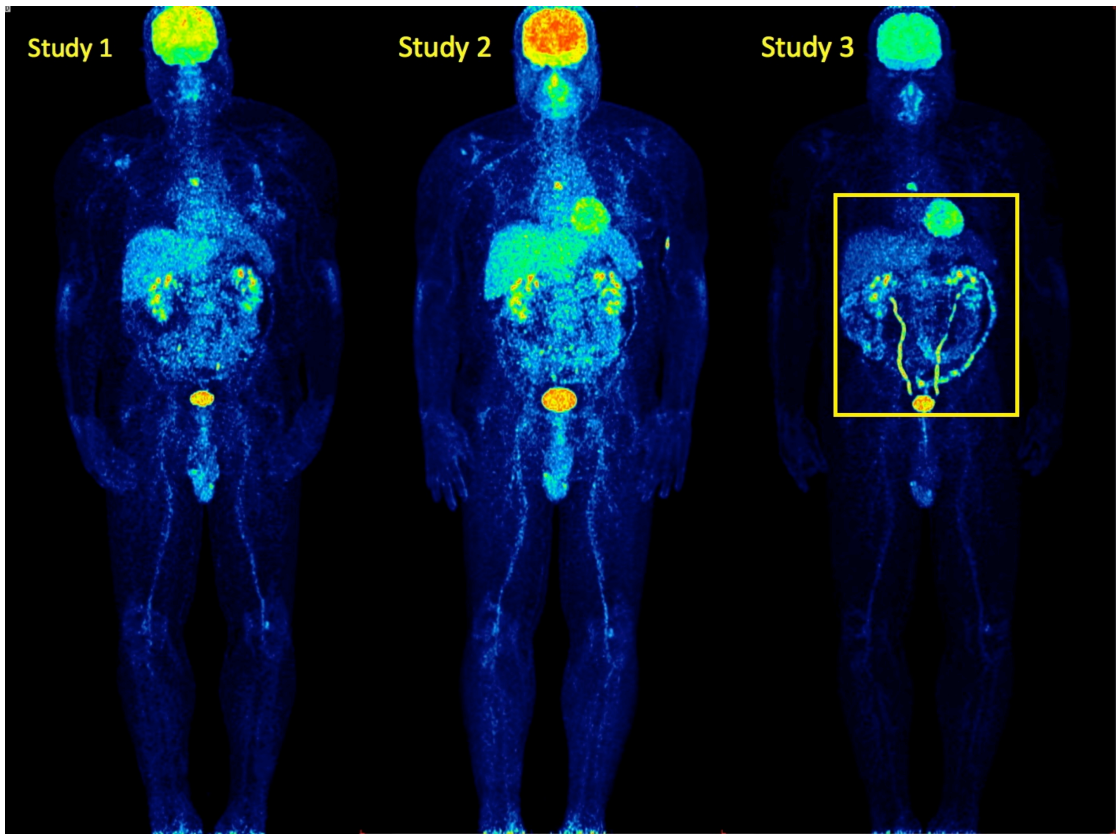

*Fig. 4.5.*

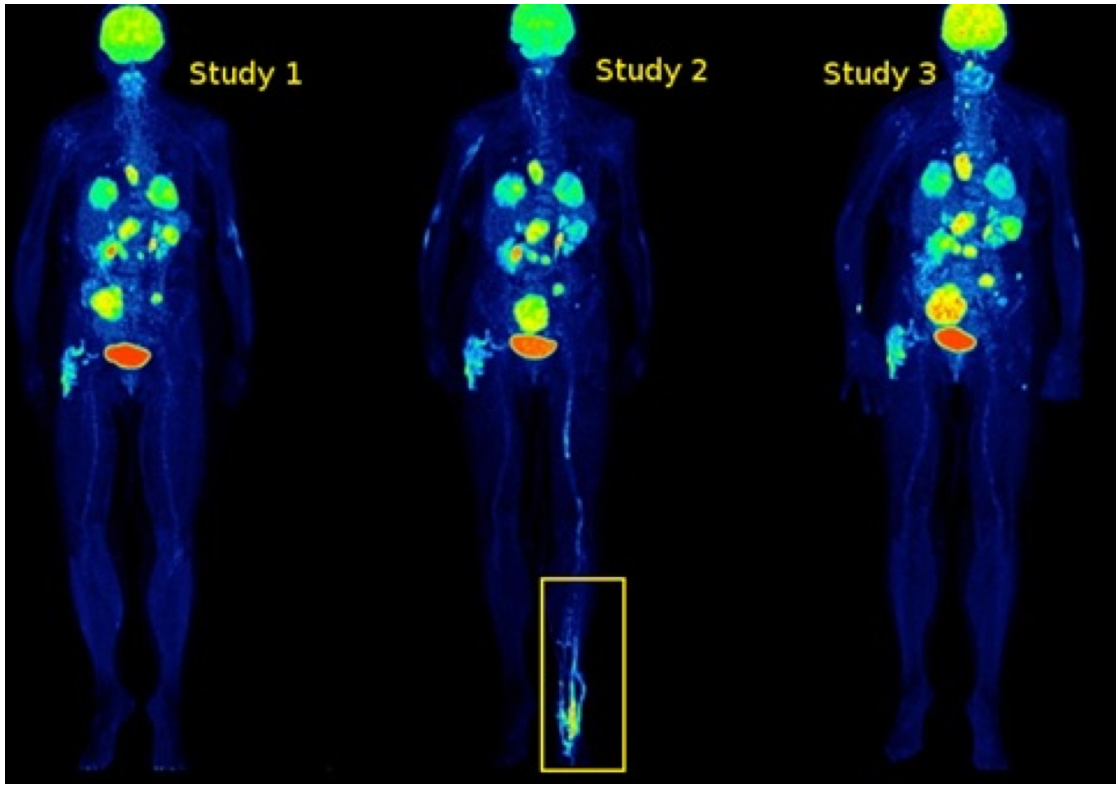

*Fig. 4.6.*

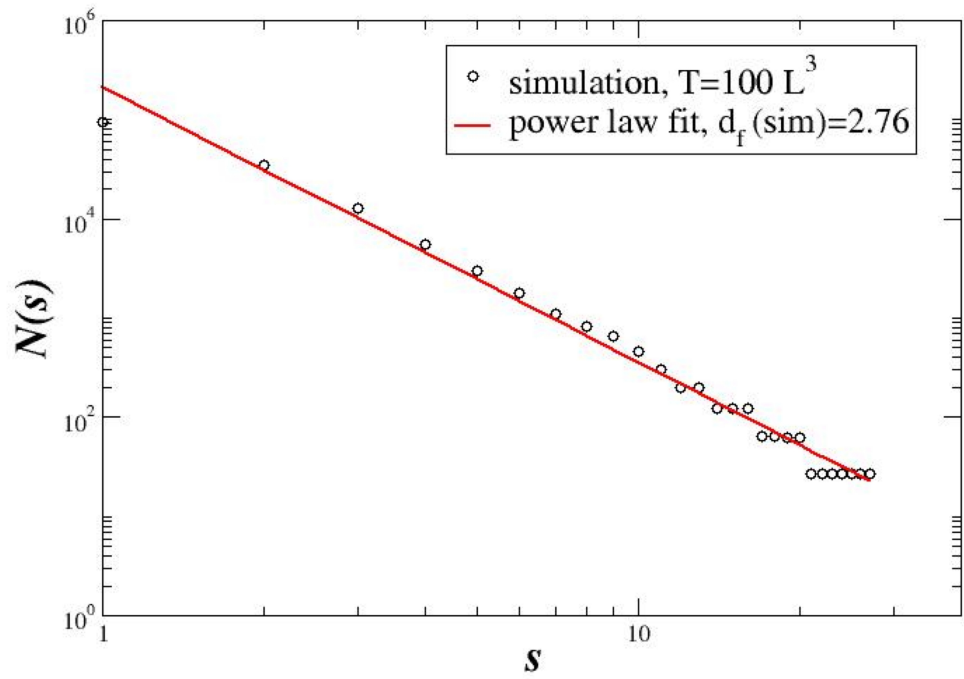

*Fig. A.1.*

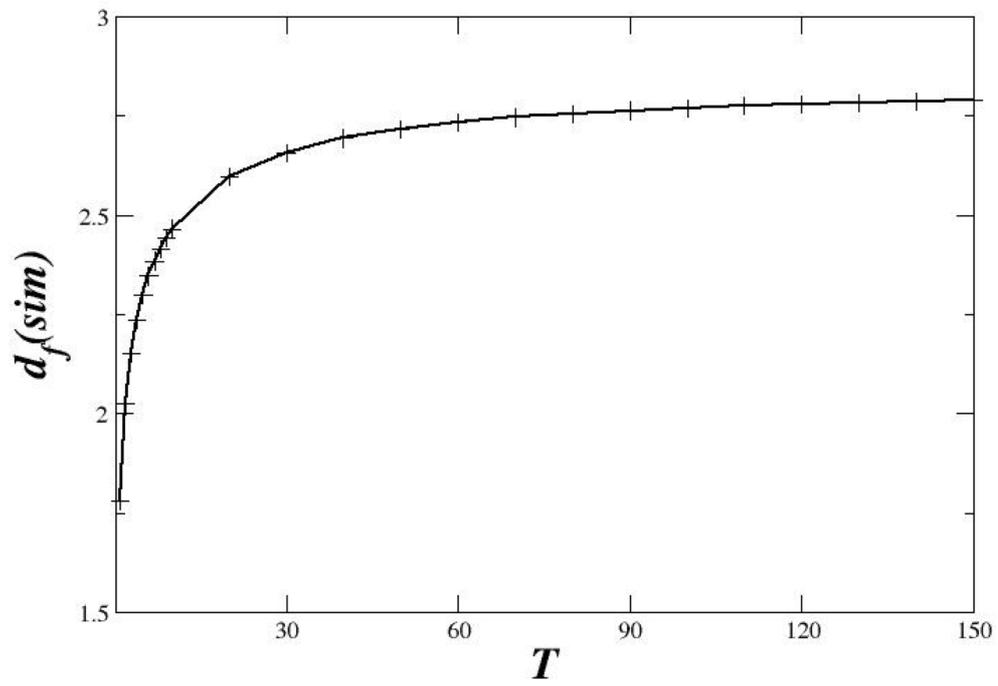

*Fig. A.2.*

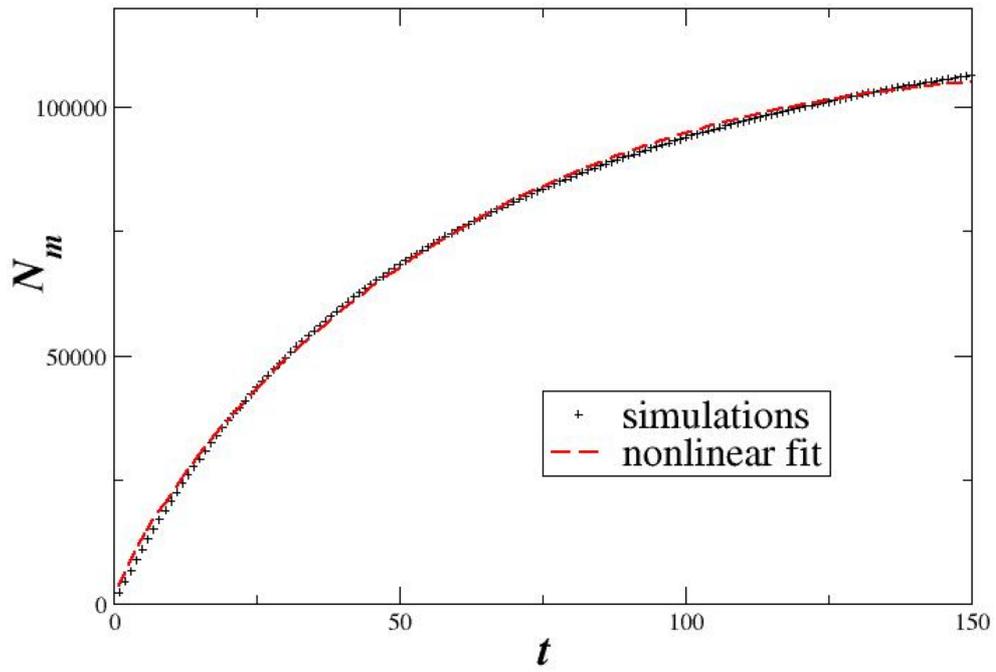

*Fig. A.3.*